\documentclass[aps,prb,groupedaddress,twocolumn,showpacs]{revtex4-1}
\usepackage{graphicx,psfrag,amsmath,calc,amssymb, comment, color,dcolumn}

\begin{document}

\title{Tighter upper bounds on the critical temperature of two-dimensional \\
  superconductors and superfluids from the BCS to the Bose regime}

\author{Tingting Shi$^{1,2}$}
\author{Wei Zhang$^{1}$}
\email{wzhangl@ruc.edu.cn}
\author{C. A. R. S{\' a} de Melo$^{2}$}
\email{carlos.sademelo@physics.gatech.edu}
\affiliation{$^{1}$ Department of Physics, Renmin University of China, Beijing 100872, China}
\affiliation{$^{2}$ School of Physics, Georgia Institute of Technology,
Atlanta, Georgia 30332, USA}

\date{\today}

\begin{abstract}
  We discuss standard and tighter upper bounds on
  the critical temperature $T_c$ of two-dimensional superconductors
  and superfluids versus particle density $n$ or filling factor $\nu$
  for continuum and lattice systems from the Bardeen-Cooper-Schrieffer (BCS)
  to the Bose regime.
  We discuss only one-band Hamiltonians, where the transition from
  the normal to the superconducting
  (superfluid) phase is governed by the
  Berezinskii-Kosterlitz-Thouless (BKT) mechanism of vortex-antivortex binding,
  such that a direct relation between the superfluid density tensor and $T_c$ exists.
  The standard critical temperature upper bound
 $T_c^{\rm up1}$ is obtained from the Ferrell-Glover-Tinkham (FGT) sum rule for
  the optical conductivity, which constrains the superfluid density tensor components.
  For the continuum,
  we show that $T_c^{\rm up1}$ is  useful only in the limit of low particle density,
  where it may be close to the critical temperature supremum $T_c^{\rm sup}$.
  For intermediate and high particle densities, 
  $T_c^{\rm up1}$ is far beyond $T_c^{\rm sup}$ for any given interaction strength.
  For the lattice, $T_c^{\rm up1}$ is useful only in the limit of low or high filling factors,
  but fails miserably at intermediate values of $\nu$.
  We demonstrate that it is imperative to consider at least the full effect of
  phase fluctuations of the order parameter for superconductivity (superfluidity)
  to establish tighter bounds over a wide range of densities or filling factors.
  Using the renormalization group, we obtain the phase-fluctuation critical temperature
  $T_c^{\theta}$ and show that it is a much tighter upper bound
  to $T_c^{\rm sup}$ than $T_c^{\rm up1}$ for all
  particle densities and filling factors throughout the BCS-Bose evolution. To achieve this,
  we go beyond textbook phase-fluctuation theories, which apply only to the BCS regime,
  where the chemical potential is essentially pinned to the Fermi energy, thus failling to describe
  the BCS-Bose crossover and the approach to the Bose limit. In sharp contrast, we show that the
  chemical potential renormalization, the order parameter equation, and
  the Nelson-Kosterlitz relation need to be solved self-consistently and simultaneously with the
  renormalization group flow equation to produce $T_c^{\theta}$ as a function of
  density or filling factor over the entire BCS-Bose crossover.
  We note that an analytic theory including modulus fluctuations of the order
  parameter valid throughout the BCS-Bose evolution is still lacking, but the inclusion of
  modulus fluctuations can only produce a critical temperature that is
  lower than $T_c^{\theta}$ and thus produce an even tighter bound to $T_c^{\rm sup}$.
 We conclude by indicating that if the measured critical temperature exceeds $T_c^{\theta}$
 in experiments involving two-dimensional single-band systems, then a non-BKT mechanism
 must be invoked to describe the superconducting (superfluid) transition.
\end{abstract}
\maketitle

\section{Introduction}
\label{sec:introdution}

Several recent experiments have studied the critical temperature $T_c$ of 
two-dimensional (2D) superconductors as a function of carrier density $n$ or
filling factor $\nu$ for various materials, including double- and
triple-layered twisted graphene~\cite{herrero-2018, herrero-2021}, lithium-intercalated
nitrides~\cite{iwasa-2018, iwasa-2021} and sulphur-doped iron
selenide~\cite{shibauchi-2021}. In all these 2D systems, the
authors~\cite{herrero-2018, herrero-2021, iwasa-2018, iwasa-2021, shibauchi-2021}
describe their results as evolving from the Bardeen-Cooper-Schrieffer (BCS) to the Bose regime
as $n$ or $\nu$ are changed, and have raised the issue
of the existence of an upper bound on $T_c$. For one-band 2D systems with
parabolic dispersion, the standard upper bound is known
to be~\cite{botelho-2006, sharapov-1999}
$T_c^{\rm up1} = \varepsilon_F/8$, in units where $k_B = 1$,
with $\varepsilon_F$ being the Fermi energy.
Extensions of this result have been proposed to flat band and
multiband systems~\cite{randeria-2019}. However, the validity of such extensions 
has been questioned in recent work showing several counter examples 
where upper bounds are arbitrarily exceeded~\cite{kivelson-2021}.

The question of the existence of an upper bound for $T_c$ in superconductors and
superfluids is of fundamental importance~\cite{kivelson-2018, kivelson-1995}.
Understanding the conditions under which such bounds exist for various systems
is key to paving the way to designing materials where room temperature
superconductivity can be achieved at ambient pressure, as suggested
by measurements of the penetration
depth of a variety of materials~\cite{uemura-1989, uemura-1991}.
However, upper bounds are practically useless if they are too far above the
supremum (least upper bound)~\cite{footnote-0},
thus when these upper bounds exist, it is essential to establish if they are tight,
that is, if they are close to the supremum $T_c^{\rm sup}$.
Identifying tighter upper bounds to $T_c^{\rm sup}$ is a much more
difficult task than merely determining a standard upper bound based on the
kinetic energy~\cite{randeria-2019}, nevertheless 
this is precisely what we propose to describe next.

Here, we study two examples where tighter upper bounds on $T_c$ 
can be established for one-band 2D systems: the continuum limit with parabolic
dispersion, and the square lattice case with cosinusoidal dispersion.
For both systems, we consider non-retarded interactions with finite range, but
we do not investigate the effects of disorder.
We show that the standard upper bound $T_c^{\rm up1}$, obtained via the bare superfluid density
$\rho_s$~\cite{botelho-2006, randeria-2019} and independent of interactions
or symmetry of the order parameter, is practically useless
away from the regime of ultralow carrier density $n$ in the continuum or away from the
limits of $\nu \to 0$ or $\nu \to 2$ in the square lattice, because it severely overestimates
the supremum $T_c^{\rm sup}$.
To remedy this issue, we demonstrate 
that much tighter bounds can be obtained by investigating the renormalized
superfluid density $\rho_s^R$ rather than the bare superfluid density $\rho_s$,
since $\rho_s^R \le \rho_s$ strictly holds. This relation arises physically because $\rho_s$ is
calculated in linear response theory, which does not include
the existence of vortices and antivortices in the superconductor or superfluid.
Large transverse current fluctuations,  due to vortices
and antivortices with quantized circulations,
screen the bare $\rho_s$ and renormalize it to $\rho_s^R$.

We obtain the phase-fluctuation critical temperature
$T_c^\theta = \pi \rho_s^R/2$ as a function of $n$ or $\nu$ 
throughout the entire BCS-Bose crossover by going beyond the standard
textbook phase-fluctuation theory for BCS superfluids and superconductors.
We obtain a set of self-consistent equations involving the order parameter, chemical potential,
and renormalized superfluid density that must be solved simultaneously with
the renormalization flow relation. Using these equations we establish
that $T_c^\theta$ is always lower than the standard upper bound $T_c^{\rm up1}$
throughout
the BCS-Bose evolution. Thus, $T_c^\theta$ is a tighter upper bound to the
supremum $T_c^{\rm sup}$ than the standard upper bound $T_c^{\rm up1}$, that is,
$T_c^{\rm sup} \le T_c^\theta \le T_c^{\rm up1}$.

We also emphasize that tighter upper bounds for $T_c$, based on $T_c^\theta$,
rely on the idea that the transition
from the superconductor or superfluid to the normal state is driven by the
Berezinskii-Kosterlitz-Thouless (BKT)~\cite{berezinskii-1970, kosterlitz-thouless-1972}
vortex-antivortex unbinding mechanism. Finally, we conclude that if an experimental $T_c$
exceeds the phase-fluctuation critical temperature $T_c^\theta$ for single-band systems,
then a non-BKT mechanism for superconductivity and superfluidity must be
invoked.

The remainder of the manuscript is organized as follows.
In Sec.~\ref{sec:hamiltonian-and-effective-action}, we describe one-band continuum and
lattice Hamiltonians and their corresponding phase-only actions.
In Sec.~\ref{sec:critical-temperature}, we obtain the self-consistency equations for the
order parameter, chemical potential, and critical temperature
$T_c^{\theta}$ that need to be solved
simultaneously with the renormalization group flow equation
for the phase-fluctuation action.
We note that this set of equations describes $T_c^{\theta}$ not
only in the standard textbook
case of the BCS regime, where the chemical potential is essentially pinned to the Fermi energy, 
but also provides a good description of the entire BCS-Bose crossover, since it includes
the renormalization of the chemical potential via the equation of state, that is,
the number equation.
We also  present a table with various definitions of upper bounds to the critical
temperature supremum $T_c^{\rm sup}$ to facilitate navigation throughout the text
and to allow for comparisons between different bounds in each of the figures shown.
In Sec. \ref{sec:results}, we present results for the phase-fluctuation critical
temperature $T_c^{\theta}$ versus density or filling factor, and show that it is a much tighter
bound to the critical temperature supremum $T_c^{\rm sup}$ than the standard
upper bound $T_c^{\rm up1}$ based on the kinetic energy bound to the bare superfluid density.
We also compare
$T_c^{\theta}$ to the critical temperatures obtained using the mean-field approximation
$T_c^{\rm mf}$ as well as to the critical temperature using the bare superfluid density
at zero temperature $T_c^{\rm up0}$. 
In Sec.~\ref{sec:beyond-standard-phase-fluctuations}, we emphasize that
our approach goes beyond the standard textbook phase-fluctuation theory and that it 
produces reliable results for $T_c^{\theta}$ not only in the BCS but also in the Bose regime.
Furthermore,  we discuss that the effects of modulus
fluctuations of the order parameter produces a new upper bound $T_c^{\theta {\rm mod}}$ that
must be lower than $T_c^{\theta}$ and thus closer to $T_c^{\rm sup}$.
However, currently, there is no analytical thermodynamic theory that
includes both modulus and phase fluctuations of the order parameter and produces
reliable critical temperature upper bounds $T_c^{\theta{\rm mod}}$. As a result,
we must rely on the full phase-fluctuation theory presented here and on the upper bound
$T_c^{\theta}$. For 2D superconductors, we also discuss that $T_c^{\theta}$ can be obtained from 
penetration depth measurements and compared to experimental $T_c$ measured resistively.
Within the BKT mechanism, we must have $T_c \le T_c^{\theta}$, however if $T_c > T_c^\theta$,
then a non-BKT mechanism must be invoked.
Lastly, in Sec.~\ref{sec:conclusions}, we conclude that $T_c^{\theta}$
is a much better upper bound to $T_c^{\rm sup}$ than the two conventional upper bounds
$T_c^{\rm up0}$ and $T_c^{\rm up1}$ for one-band two-dimensional superconductors
and superfluids.

\section{Hamiltonian and Effective Action}
\label{sec:hamiltonian-and-effective-action}

To construct tighter upper bounds on $T_c$,
we discuss 2D continuum and lattice Hamiltonians and derive their corresponding phase-fluctuation
actions below.

In the continuum, we start from the Hamiltonian density 
\begin{equation}
{\cal H} ({\bf r})
=
{\cal H}_{\rm K} ({\bf r})
+
{\cal H}_{\rm I} ({\bf r}), 
\end{equation}
for a single band system in units where $\hbar=k_B=1$. The
kinetic energy density is
\begin{equation}
{\cal H}_{\rm K} ({\bf r})
=
\sum_s
\psi^\dagger_{s} ({\bf r}) \left[  -\frac{\nabla^2}{2m} \right] \psi_s ({\bf r}),
\end{equation}
and the interaction Hamiltonian density is 
\begin{equation}
{\cal H}_{\rm I} ({\bf r})
=
\int d^2 {\bf r}^\prime
V ({\bf r}, {\bf r}^\prime)
\psi_{\uparrow}^\dagger ({\bf r}) \psi_{\downarrow}^\dagger ({\bf r}^\prime)
\psi_{\downarrow} ({\bf r}^\prime)  \psi_{\uparrow} ({\bf r}),
\end{equation}
with
$V ({\bf r}, {\bf r}^\prime) = - V_s g (\vert {\bf r} - {\bf r}^\prime\vert/ R)$.
The  magnitude of the $s$-wave attractive interaction $V_s$ has units of energy,
and the dimensionless function $g (\vert {\bf r} - {\bf r}^\prime\vert/ R)$
has  spatial range $R$.
Fermions with spin projection $s$ at position ${\bf r}$ are represented by 
the field operator $\psi_s^{\dagger} ({\bf r})$.

In the square lattice, we start from an extended Fermi-Hubbard Hamiltonian 
\begin{equation}
\label{eqn:extended-attractive-Hubbard-model}  
H =
-t \sum_{\langle ij \rangle, s}  \psi_{i s}^\dagger  \psi_{j s}
-
U \sum_{i s} {\hat n}_{i \uparrow} {\hat n}_{i \downarrow}
-
\sum_{i < j, s s^\prime} V_{ij} {\hat n}_{i s} {\hat n}_{j s^\prime},
\end{equation}
where ${\hat n}_{i s} = \psi_{i s}^\dagger \psi_{i s}$ is the fermion number
operator at site $i$ with spin $s$. The nearest neighbor hopping is $t$,
the local (on-site) interaction is $U > 0$,
and the interaction between fermions in sites $i$ and $j$ is $V_{ij} > 0$.
Both the on-site and nearest neighbor interactions are attractive.

For both the continuum and lattice cases, the Hamiltonians in momentum space
can be written as
\begin{equation}
\label{eqn:hamiltonian-momentum-space}
H =
 \sum_{{\bf k} s} 
\varepsilon_{\bf k} \psi_{{\bf k}, s}^\dagger \psi_{{\bf k}, s} + 
\sum_{{\bf k} {\bf k}^\prime {\bf q}} V_{{\bf k}{\bf k}^\prime}
b_{{\bf k}{\bf q}}^\dagger b_{{\bf k}^\prime {\bf q}} ,
\end{equation}
where 
$b_{{\bf k}{\bf q}} = \psi_{-{\bf k}+{\bf q}/2, \downarrow} \psi_{{\bf k}+{\bf q}/2, \uparrow}$
is the pairing operator, 
$\varepsilon_{\bf k} = {\bf k}^2/2m$ for the continuum, 
and $ \varepsilon_{\bf k} = - 2t \left[ \cos (k_x a) + \cos (k_y a) \right] $ for the square lattice.
The summation over ${\bf k}$ and ${\bf k}^{\prime}$ represent integrals
in the continuum and discrete sums in the lattice. The momentum-space interaction
$V_{{\bf k} {\bf k}^\prime}$ is the double Fourier transform of the real space interactions.

In the examples discussed below, we focus only on $s$-wave pairing,
and thus, we will use an expansion of 
the momentum space interaction $V_{{\bf k} {\bf k}^\prime}$ in terms of its continuum
or lattice angular momentum~\cite{duncan-2000, annett-1990, iskin-2005}
and keep only the $s$-wave component. In this case,
the interaction potential can be approximated by the separable form
\begin{equation}
\label{potential}
V_{{\bf k}{\bf k}'} = -V_s 
\Gamma_s ({\bf k}) \Gamma_s({\bf k}^{\prime}) .
\end{equation}
In the continuum, for an attractive well with depth $V_s$ and radius $R$,
the symmetry factor can be approximated by 
$
\Gamma_s ({\bf k}) = 
\left( 1 + k/k_R \right)^{-1/2},
$
where $k_R \sim  R^{-1}$ plays the role of the interaction range in momentum
space and $k$ is the modulus of ${\bf k}$, that is, $k = \vert {\bf k} \vert$.
This parametrization is necessary to produce a
separable potential in momentum space that behaves both at small and large momenta
in a form that is compatible to an acceptable real space interaction
potential $V({\bf r}, {\bf r}^{\prime})$. 

In the square lattice, which has $C_4$ point group, the
symmetry factor for conventional $s$-wave pairing is $\Gamma_s ({\bf k}) = 1$,
when, for instance,  only local attractive interactions are considered,
while the symetry factor for extended $s$-wave pairing is
$\Gamma_s ({\bf k}) = \cos (k_x a) + \cos (k_y a)$, when, for example, nearest neighbor
attractive interactions are included.
For more general lattices, $\Gamma_s ({\bf k})$ are given by
the irreducible representations of the point group of the lattice~\cite{annett-1990}
compatible with $s$-wave symmetry. 

We introduce the chemical potential $\mu$ and  the order parameter
$\Delta$ for superfluidity in terms of its modulus $\vert \Delta \vert$ and phase $\theta$.
By fixing the modulus $\vert \Delta \vert$ and including full phase fluctutations, 
the effective action is
\begin{equation}
\label{eqn:effective-action}
S_{\rm eff} =
S_{\rm sp} \left( \vert \Delta \vert \right)
+ 
S_{\rm ph} ( \vert \Delta \vert, \theta).
\end{equation}
The first contribution to $S_{\rm eff}$ is the 
saddle-point action 
\begin{equation}
\label{eqn:saddle-point-action}
  S_{\rm sp} 
=
\sum_{\bf k}
\left[
\frac{\left( \xi_{\bf k} - E_{\bf k} \right)}{T}
-
2 \ln \left(  1 + e^{-E_{\bf k}/T}\right)
\right] + \frac{\vert \Delta \vert^2}{T V_s},
\end{equation}
where
$
E_{\bf k} = \sqrt{\xi_{\bf k}^2 + |\Delta_{\bf k}|^2}
$
is the energy of quasiparticles with $\xi_{\bf k} = \varepsilon_{\bf k} - \mu$, and 
$\vert \Delta_{\bf k}\vert  = \vert \Delta \vert  \vert \Gamma_{s} ({\bf k}) \vert$
is the modulus of order parameter function for $s$-wave pairing.
The second contribution
\begin{equation}
\label{eqn:phase-fluctuation-action}
S_{\rm ph} 
= \frac {1} {2} \int dr 
\left\{
\sum_{ij} \rho_{ij} \, \partial_i \theta(r) \partial_j \theta(r)
+
\kappa_s \left[ \partial_{\tau} \theta (r) \right]^2 
\right\}
\end{equation}
represents the phase-fluctuation action 
in the low frequency and long wavelength limit.
The integrals run over position ${\bf r}$ and imaginary time $\tau$, that is,
$r = ({\bf r}, \tau)$, with
$\int dr \equiv \int_0^{1/T} d\tau \int d^2{\bf r}$.
The first term in Eq.~(\ref{eqn:phase-fluctuation-action}),
\begin{equation}
\label{eqn:superfluid-density}
\rho_{ij} 
=
\frac {1} {4 L^2} \sum_{\bf k}
\left[
2 n_{\rm sp} ({\bf k}) \partial_i \partial_j \xi_{\bf k} - 
Y_{\bf k} \partial_i \xi_{\bf k} \partial_j \xi_{\bf k}
\right] ,
\end{equation}
is the superfluid density tensor,  where $\partial_i$ is the partial 
derivative with respect to momentum $k_i$ with $i = \{ x, y \}$.
The first term in Eq.~(\ref{eqn:superfluid-density}),
\begin{equation}
\label{eqn:momentum-distribution}  
n_{\rm sp} ({\bf k}) = \frac {1} {2} \left[
1 - (\xi_{\bf k}/ E_{\bf k}) 
\tanh \left(  E_{\bf k} / 2 T \right) \right]
\end{equation}
is the momentum distribution per spin state, while the second, 
$
Y_{\bf k} = (2T)^{-1} 
{\rm sech}^2 (E_{\bf k} / 2T)
$
is the Yoshida function.
The superfluid density tensor is diagonal
$\rho_{ij} = \rho_s \delta_{ij}$, where
\begin{equation}
\rho_s = (1/4mL^2) \sum_{\bf k} \left[ 2 n_{\rm sp} ({\bf k}) - (k_x^2/m) Y_{\bf k}\right]
\end{equation}
is the superfluid density for the continuum and
\begin{equation}
\rho_s =
(ta^2/L^2) \sum_{\bf k} \left[  \cos (k_x a) n_{\rm sp}({\bf k}) - t \sin^2 (k_x a) Y_{\bf k} \right]
\end{equation}
for the square lattice.
The second term in Eq.~(\ref{eqn:phase-fluctuation-action}) is
\begin{equation}
\label{eqn:compressibiity}
\kappa_{s} 
=
\frac {1} {4 L^2} \sum_{\bf k} 
\left[
\frac {|\Delta_{\bf k}|^2} { E_{\bf k}^3 }
\tanh \left( \frac { E_{\bf k} }  { 2 T}  \right) + 
\frac { \xi_{\bf k}^2 } { E_{\bf k}^2 } Y_{\bf k}
\right] ,
\end{equation}
with $\kappa_s = \kappa/4$, where
$\kappa = \partial n/ \partial \mu \vert_{T,V}$ is related to the
thermodynamic compressibility ${\cal K} = \kappa/n^2$.

The phase of the order parameter can be separated as 
$\theta ({\bf r}, \tau) = \theta_c ({\bf r}, \tau) + \theta_v ({\bf r}) $,
where the $\tau$-dependent (quantum) term
$\theta_c ({\bf r}, \tau)$ is due to collective modes (longitudinal velocities),
and the $\tau$-independent (classical) term $\theta_v ({\bf r})$ is due to vortices 
(transverse velocities). This leads to the action
\begin{equation}
S_{\rm ph} = S_{c} + S_{v},
\end{equation}
since the longitudinal and transverse velocities are orthogonal.
The collective mode action is 
\begin{equation}
S_{c} = \frac{1}{2} \int dr 
\left[
\rho_s \left[ \nabla \theta_{c} (r) \right]^2
+
\kappa_{s} \left[ \partial_\tau \theta_{c} (r) \right]^2
\right],
\end{equation}
while the vortex action is  
\begin{equation}
\label{eqn:vortex-action}
  S_{v}
=
\frac{1}{2T} \int d^2{\bf r} \,
\rho_s\left[ \nabla \theta_v ({\bf r}) \right]^2.
\end{equation}
The vortex contribution arises from the transverse velocity
${\bf v}_t = \nabla \theta_v ({\bf r})$, where $\nabla \cdot {\bf v}_t ({\bf r}) = 0$.
Using the relation
$
\nabla \times {\bf v}_t ({\bf r})
=
2\pi {\hat {\bf z}} n_v ({\bf r}),
$
where
$
n_v ({\bf r})
=
\sum_{i} n_i \delta ({\bf r} - {\bf r}_i ),
$
is the vortex density and $n_i = \pm 1$ is the vortex topological charge (vorticity)
at ${\bf r}_i$, we write
\begin{equation}
\label{eqn:vortex-action-and-interaction-potential}  
S_v =
\frac{\rho_s}{2T}
(2\pi)^2 \int\int d^2{\bf r} d^2{\bf r}^{\prime}
n_v ({\bf r}) G ({\bf r} - {\bf r}^{\prime}) n_v ({\bf r}^{\prime}),
\end{equation}
where $G ({\bf r}_i - {\bf r}_j)$ is the dimensionless interaction potential between
the vortices $n_v ({\bf r})$ and $n_v ({\bf r}^{\prime})$,
satisfying Poisson's equation $\nabla^2_{\bf r} G ({\bf r} - {\bf r}^\prime) = 0.$ 
The formal solution to this differential equation is
$
G ({\bf r} - {\bf r}^{\prime}) =
\int d{\bf k}
e^{i {\bf k} \cdot ({\bf r} - {\bf r}^{\prime})}
/
\left[ (2\pi)^2 \vert {\bf k} \vert^2 \right],
$
with lower $(\pi/L)$ and upper $(\pi/d)$ cutoffs to regularize the action. Here,  $L$ is
the linear system size and $d$ is the minimum separation between vortices, leading
to
$
G ({\bf r} - {\bf r}^{\prime})
=
\left[ \ln (L/d) - \ln ( \vert {\bf r} - {\bf r}^{\prime} \vert / d) + C \right]/2\pi,
$
where $C$ is a constant. In the thermodynamic limit, the system size $L \to \infty$,
and topological charge neutrality $\sum_{i} n_i = 0$ leads to
\begin{equation}
\label{eqn:vortex-action-core-energy}
S_{v}
=
- 2\pi \frac{\rho_s}{2T} \sum_{i \ne j} n_i n_j
\ln \left[ \frac{\vert {\bf r} - {\bf r}^{\prime} \vert}{d} \right]
+ \sum_i \frac{E_c}{T} n_i^2 ,
\end{equation}
where $E_c$ is the vortex core energy.  Note that $E_c$ must be proportional to the bare
superfluid density $\rho_s$ since the phase-only fluctuation theory describing
vortices has one energy scale, which is controlled by $\rho_s$, as explicitly seen
in the vortex action shown in Eq.~(\ref{eqn:vortex-action}).

\section{Critical Temperature}
\label{sec:critical-temperature}

We are interested in calculating the phase-fluctuation critical
temperature $T_c^{\theta}$ and in using it as a better upper bound for the critical temperature
supremum $T_c^{\rm sup}$ of two-dimensional superconductors and superfluids when
density or filling factor are varied during the BCS-Bose evolution.
For this purpose, it is absolutely necessary
to go beyond the standard textbook analysis of the phase-fluctuation
action that neglects the chemical potential renormalization when
density or filling factor are changed for fixed interactions.
We show below, that the renormalization of the chemical potential by
phase fluctuations is essential to recover a physically acceptable
description of the BCS-Bose crossover and the Bose limit.
The definition of various temperatures that are used below
are summarized in Table~\ref{tab:tc}.

\begin{table}
\caption{\label{tab:tc}List of temperatures obtained by different methods.}
\begin{ruledtabular}
\begin{tabular}{ll}
 $T_c$  & Description \\
\hline
$T_c^{\rm sup}$ & Critical temperature supremum \\
$T_c^\theta$ & Phase-fluctuation critical temperature\\ 
$T_c^{\rm up0}$ & Upper bound set by superfluid density $\rho_s(T=0)$\\ 
$T_c^{\rm up1}$ & Upper bound set by kinetic energy\\
$T_c^{\rm mf}$ & Mean-field critical temperature\\ 
$T_c^{\theta {\rm mod}}$ & Modulus and phase-fluctuation $T_c$
\end{tabular}
\end{ruledtabular}
\end{table}

The determination of $T_c^{\theta}$ requires self-consistency relations for $\vert \Delta \vert$
and $\mu$ for a given temperature $T$ and the renormalization group flow equation for the
BKT mechanism. These self-consistency equations  are obtained from 
effective action $S_{\rm eff} = S_{\rm sp} + S_{\rm ph}$, shown in
Eqs.~(\ref{eqn:effective-action}),~(\ref{eqn:saddle-point-action}),
and ~(\ref{eqn:phase-fluctuation-action}), as follows.
The order parameter equation is obtained through the stationarity condition 
$
\delta S_{\rm sp} / 
\delta \Delta^{*} = 0 ,
$
leading to 
\begin{equation}
\label{eqn:order-parameter}
\frac {1} {V_{s}}  = 
\sum_{\bf k} \frac {|\Gamma_{s} ({\bf k})|^2} {2 E_{\bf k}}
\tanh \left( \frac {E_{\bf k}} {2 T} \right) .
\end{equation}

To relate the chemical potential $\mu$ and the particle density $n = N/L^2$,
where $N$ is the total number of particles per band and $L^2$ is the area of the sample,
we use the thermodynamic relation
$
n =
-\partial {\widetilde \Omega}/ \partial\mu \vert_{T, V},
$
with ${\widetilde  \Omega} = \Omega/L^2$, where
\begin{equation}
\label{eqn:thermodynamic-potential}  
\Omega= \Omega_{\rm sp} + \Omega_{\rm ph}
\end{equation}
is the thermodynamic potential.
The first term is the saddle point contribution
\begin{equation}
  \Omega_{\rm sp}= T S_{\rm sp},
\end{equation}
where $S_{\rm sp}$ is given in
Eq.~(\ref{eqn:saddle-point-action})
and the second term is due to phase fluctuations
\begin{equation}
  \Omega_{\rm ph} = \Omega_{v} + \Omega_c.
\end{equation}
The contribution due to vortices is $\Omega_{v} = -T \ln {\cal Z}_v$,
where ${\cal Z}_v = \int d\theta_v e^{-S_v}$, arises from transverse phase
fluctuations $\theta_v ({\bf r})$. The contribution due to collective modes
is
\begin{equation}
  \Omega_c = \Omega_{\rm cm} + \Omega_{\rm zp},
\end{equation}
resulting from integration over longitudinal phase fluctuations $\theta_c ({\bf r}, \tau)$. 
Here, 
\begin{equation}
\Omega_{\rm cm} = \sum_{\bf q} 
T\ln \left[ 1 - \exp \left( -  { \omega_{\bf q} / T } \right) \right]
\end{equation}
and 
$\Omega_{\rm zp} = \sum_{{\bf q}} \omega_{\bf q}/2$,
where $\omega ({\bf q}) = c |{\bf q}|$ is the frequency and 
$c = \sqrt {\rho_s/\kappa_{s}}$ is the speed of sound. 

Thus, the number equation
\begin{equation}
\label{eqn:number}
  n =  n_{\rm sp} + n_{\rm cm} + n_{\rm zp} + n_v
\end{equation}
has four contributions of the form
$
n_{j} = - \partial {\widetilde \Omega}_{j} / \partial \mu \vert_{T,V}$,
with $j = \{{\rm sp}, {\rm cm}, {\rm zp}, v \}$. The last three contributions
are due to phase-fluctuations, that is, $n_{\rm ph} = n_{\rm cm} + n_{\rm zp} + n_v$.
Here the saddle point term is
$n_{\rm sp}= 2 \sum_{\bf k} n_{\rm sp} ({\bf k})$, where $n_{\rm sp} ({\bf k})$ is given
in Eq.~(\ref{eqn:momentum-distribution}), while $n_{\rm cm}$, $n_{\rm zp}$ and $n_v$
are obtained from their respective
${\widetilde\Omega}_j$. We emphasize that the
number equation, shown in Eq.~(\ref{eqn:number}),
plays an important role in the renormalization of the
chemical potential $\mu$ due to phase fluctuations and transcends 
textbook phase-fluctuation theories that fix the chemical potential to the Fermi energy,
and thus can only be applied in the BCS regime.
In Sec.~\ref{sec:beyond-standard-phase-fluctuations},
we show a specific example where the standard textbook results fail miserably
in strong coupling,
but this is also the case at low densities in the continuum and for
filling factors $\nu \approx 0$ or
$\nu \approx 2$ for one-band systems. In these cases, interaction terms are so strong in
comparison to the kinetic energy that any BCS-like approach is doomed to fail.

Within the BKT mechanism, the phase-fluctuation critical temperature  $T_c^{\theta}$
is obtained from the vortex thermodynamic potential $\Omega_v$ associated
with the vortex action
in Eqs.~(\ref{eqn:vortex-action}) or~(\ref{eqn:vortex-action-core-energy}),
and is given by the Nelson-Kosterlitz~\cite{nelson-kosterlitz-1977} relation 
\begin{equation}
\label{eqn:critical-temperature}
T_{c}^{\theta}
= \frac {\pi} {2} \rho_s^{R}  ( \mu, \vert \Delta \vert, T_c^{\theta} ),
\end{equation}
as $T \to T_c^{\theta}$ from below, 
where 
\begin{equation}
\label{eqn:renormalized-superfluid-density} 
\rho_s^{R}
=
\rho_s -
\frac{\rho_s^2}{2T}
\lim_{{\bf q} \to {\bf 0}}
\frac{\langle n_v ({\bf q}) n_v (-{\bf q}) \rangle}{\vert {\bf q} \vert^2}, 
\end{equation}
is the renormalized superfluid density.
Here, $n_v ({\bf q})$ is the Fourier transform of vortex density
$n_v ({\bf r})$. Notice that $\rho_s^{R}$ is a function of 
chemical potential $\mu$, order parameter modulus $\vert \Delta \vert$ and
temperature $T$, and thus it is sensitive to the renormalization of the chemical
potential as temperature or density are changed for a given interaction strength.

From Eq.~(\ref{eqn:renormalized-superfluid-density}), it is clear
that $\rho_s^{R} \le \rho_s$ at any temperature $T$, because the correlation
function
$
{\cal F}
=
\lim_{{\bf q } \to {\bf 0}} \langle  ns_v ({\bf q}) n_v (- {\bf q}) \rangle/\vert {\bf q} \vert^2
$
is strictly non-negative, that is, ${\cal F} \ge 0$.
This implies that the upper bound $T_c^{\rm up1}$ based
on the bare superfluid density $\rho_s$~\cite{botelho-2006, randeria-2019} is not tight, and
therefore may severely overestimate the least upper bound, that is,
the supremum $T_c^{\rm sup}$. The phase-fluctuation critical temperature $T_c^{\theta}$, given in 
Eq.~(\ref{eqn:critical-temperature}), must be viewed as a tighter upper bound to
the supremum $T_c^{\rm sup}$.

To establish $T_c^{\theta}$, we combine
Eqs.~(\ref{eqn:vortex-action-core-energy}) and~(\ref{eqn:renormalized-superfluid-density})
to obtain the solution of the renormalization group
flow equations~\cite{kosterlitz-1974, nelson-1977, chaikin-1995} leading to 
\begin{equation}
\label{eqn:RG-solution}
y^2(l) - \frac{1}{2\pi^3}
\left[
\frac{2}{K(l)} + \pi {\rm ln} K(l)
\right] =  A,
\end{equation}
for running variables $K(l)$ and $y(l)$.  The initial conditions are $K(0) = \rho_s/T$
and $y(0) = {\rm exp}(-E_c/T)$, satisfying  the relation
$y(0) = {\rm exp}{\left[ -E_c K (0)/\rho_s\right] }$. The flow of $\rho_s$
is described by $K (\ell) $, and
the flow of the vortex fugacity is represented by $y (\ell)$. 
The constant $A$ determines the family of flow curves in the $K^{-1}$-$y$ plane.
We stress that our approach goes beyond the standard textbook description
of phase-fluctuations
which is only valid in the BCS regime, where the phase-fluctuation equation of state, 
the number equation shown in Eq.~(\ref{eqn:number}),
is not included in the self-consistent solution and the chemical
potential is pinned to the Fermi energy, that is, 
$\mu \approx \varepsilon_F$.  Here, however,
the number equation and the renormalization of the chemical potential play a very
important role. For one-band systems, the chemical potential renormalization
is particularly important in two regimes: at low densities in the continuum and
at filling factors $\nu \approx 0$
and $\nu \approx 2$ in the lattice with fixed interactions;
or at intermediate and strong interactions with fixed density or filling factor.
See further discussion in Sec.~\ref{sec:beyond-standard-phase-fluctuations}.

\begin{figure}[t]
\begin{center}
\includegraphics[width=0.98\linewidth]{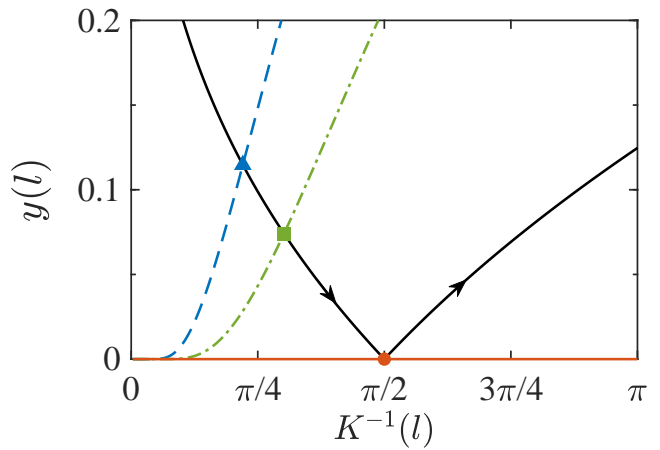}
\end{center}
\vspace{-6mm}
\caption{Critical flow line (solid black) and examples of initial conditions
$y(0) = e^{-E_c K(0)/\rho_s}$ lines for $E_c = 1.5 \rho_s$ (dashed blue),
for $E_c = \pi^2 \rho_s/4$ (dot-dashed green), and $E_c \to \infty$ (solid red).
The critical points are the solid blue triangle for $E_c = 1.5 \rho_s$,
the solid green square for $E_c = \pi^2 \rho_s/4$ (XY model), and
the solid red  circle for $E_c \to \infty$, which is also the fixed point of critical flow line.
When $E_c \to \infty$, $T_c^{\theta} \to \pi \rho_s/2$, thus for any  $E_c \le  \infty$, then
$T_c^{\theta} = \rho_s/K_c \le  \pi \rho_s/2$.
}
\label{fig:one}
\end{figure}

In Fig.~\ref{fig:one}, the critical flow curve,  for which
$A = \left[ \ln \pi/2 - 1 \right]/\pi^2 = -0.0278$, is shown.
The fixed point at $(K_*^{-1}, y_*) = (\pi/2, 0)$ leads to 
the relation $T_c^\theta/\rho_s^R = \pi/2$ in Eq.~(\ref{eqn:critical-temperature}).
From the intersection between the curve
$y(0) = {\rm exp}{\left[ -E_c K (0)/\rho_s\right] }$ and the critical flow line, we obtain the
relation $T_c^\theta/\rho_s = K_c^{-1}$, which allows us to relate $\rho_s^R$ and $\rho_s$ 
via $\rho_s^R = \rho_s (2/\pi K_c)$. As seen in Fig.~\ref{fig:one},  $K_c^{-1} \le \pi/2$,
that is, $( 2/\pi K_{c}) \le 1$ for any value
of the vortex core energy $E_c$. Therefore, $\rho_s^{R}$ is
always less or equal to $\rho_s$, that is, $\rho_s^{R} \le \rho_s$. The equality between
$\rho_s^R$ and $\rho_s$ occurs only when $E_c \to \infty$.

An important consequence of $\rho_s^R \le \rho_s$ is that
$T_c^\theta = \pi \rho_s^R/2 \le \pi \rho_s/2$. Since the supremum $T_c^{\rm sup}$ is always
upper-bounded by the phase-fluctuation critical temperature $T_c^\theta$, that is,
$T_c^{\rm sup} \le T_c^{\theta}$,
it is clear that a lower upper bound is reached using $\rho_s^R$ rather than $\rho_s$.
Thus, if the proportionality constant between $E_c$ and $\rho_s$ is known for specific
model Hamiltonians, then $\rho_s^{R}$ can be calculated and
an even tighter upper bound to $T_c^{\rm sup}$ can be obtained. 

Physically, $T_c^{\theta}$ is a better upper bound to $T_c^{\rm sup}$ than
\begin{equation}
\label{eqn:upper-bound-0}
T_c^{\rm up0} = \pi \rho_s (T = 0)/2
\end{equation}
as suggested for superconductors with small superfluid density~\cite{kivelson-1995},
where phase fluctuations are important.
Furthermore, $T_c^{\theta}$ is a better upper bound than 
\begin{equation}
\label{eqn:upper-bound-1}
T_c^{\rm up1} = \pi {\widetilde \rho}_{s_1}/2,
\end{equation}
where 
${\widetilde \rho}_{s_1}$ is an upper bound of 
\begin{equation}
\rho_{s_1} =  (1/4L^2) \sum_{\bf k} 2 n_{\rm sp} ({\bf k}) \partial_{x}^{2} \xi_{\bf k},
\end{equation}
which is the first term shown in Eq.~(\ref{eqn:superfluid-density})
when the bare superfluid density is diagonal, that is, $\rho_{ij} = \rho_s \delta_{ij}$,
and $\xi_{\bf k}$ is isotropic in the continuum
(${\rm C}_\infty$ or ${\rm SO}(2)$ symmetric)
or in the square lattice (${\rm C}_4$ symmetric).

In the continuum, the relation 
\begin{equation}
\rho_{s} (T = 0)  =  n/4m
\end{equation}
arises due to Galilean invariance~\cite{botelho-2006},
and $\rho_s (T)$, at any temperature $T$, is upper bounded
by $\rho_{s} (T = 0)$, that is, by the ratio between the maximum pair density $n/2$
and the fermion pair mass $2m$. A similar analysis based on the Ferrell-Glover-Tinkham
(FGT)~\cite{ferrell-1958, tinkham-1959, tinkham-1975} sum rule
for the optical conductivity, first derived by Kubo~\cite{kubo-1957},  requires that 
$\rho_s (T) \le n/4m$. This means that ${\widetilde \rho}_{s_1} = n/4m$ leading to the standard
upper bound~\cite{botelho-2006, randeria-2019}
\begin{equation}
\label{eqn:tc-up1-continuum}  
T_c^{\rm up1} = \pi n/8m = \varepsilon_F/8,
\end{equation}
which is also known to apply
to anisotropic superfluids~\cite{devreese-2014, devreese-2015, randeria-2019},
like those with spin-orbit coupling~\cite{devreese-2014, devreese-2015}.
Thus, in the continuum, it is clear that the two upper bounds $T_c^{\rm up0}$
and $T_c^{\rm up1}$ are exactly the same.

For the square lattice, $T_c^{\rm up0}$ is still given by Eq.~(\ref{eqn:upper-bound-0})
and $T_c^{\rm up1}$ by Eq.~(\ref{eqn:upper-bound-1}), which becomes
\begin{equation}
\rho_{s_1} = (t a^2/L^2) \sum_{\bf k} \cos (k_x a) n_{sp} ({\bf k}) \le {\widetilde \rho}_{s_1},
\end{equation}
where ${\widetilde \rho}_{s_1}=  t \nu/2$
with $\nu$ being the filling factor of the band. 
This upper bound is consistent with the FGT sum rule~\cite{hirsch-2000, chubukov-2011}.
Since the momentum distribution
$n_{\rm sp} ({\bf k}) \ge 0$ for any ${\bf k}$, the result above is easily obtained
through the successive bounds
$\rho_{s_1} = (t a^2/L^2) \sum_{\bf k} \cos (k_x a) n_{\rm sp} ({\bf k}) \le
(t a^2/L^2) \sum_{\bf k} \vert \cos (k_x a) \vert  n_{\rm sp} ({\bf k}) \le
(t a^2/L^2) \sum_{\bf k} n_{\rm sp} ({\bf k}) \le  (ta^2/L^2) (N/2) = t \nu/2$,
where we used $L^2 = N_{si} a^2$ and the filling factor $\nu = N/N_{si}$,
with $N_{si}$ being the number of sites and $N$ being the number of particles.
Using particle-hole symmetry, a similar bound
$\rho_{s_1} \le t (2 - \nu)/2$ applies, that is, ${\widetilde \rho}_{s_1} = t (2 - \nu)/2$. 
Combining these two-bounds leads to the standard upper bound
\begin{equation}
\label{eqn:tc-up1-lattice}  
T_c^{\rm up1} = t (\pi/4) {\rm min} \{ \nu, (2 - \nu) \}.
\end{equation}
Notice that $T_c^{\rm up1}$ in Eqs.~(\ref{eqn:tc-up1-continuum}) and~(\ref{eqn:tc-up1-lattice})
is determined solely by the kinetic energy, and is directly related to the density $n$
or filling factor $\nu$, respectively.

Setting the mean-field temperature $T_c^{\rm mf}$, as well as the modulus and
phase fluctuation $T_c^{\theta {\rm mod}}$ aside,
the other four temperatures defined in Table~\ref{tab:tc}
are ordered in the sequence
$T_c^{\rm sup} \le T_c^{\theta} \le T_c^{\rm up0} \le T_c^{\rm up1}$,
with $T_c^{\theta}$ being the best upper bound to $T_c^{\rm sup}$
and $T_c^{\rm up1}$ being the worst, as discussed next.

\begin{figure}[t]
\begin{center}
\includegraphics[width=0.98\linewidth]{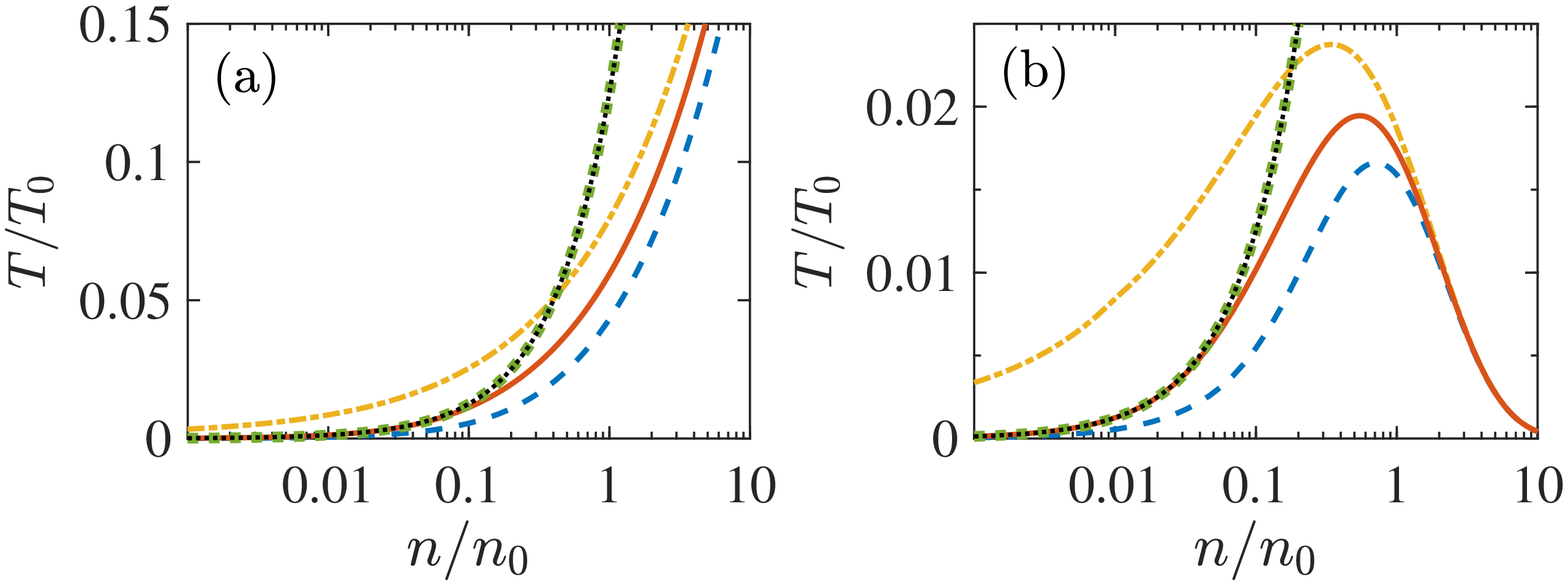}   
\end{center}
\vspace{-6mm}
\caption{Plots of $T_c$ versus $n$, in units of
$\varepsilon_0 = k_0^2/2m = T_0$ and $n_0 = k_0^2/2\pi$. 
Results for zero-ranged interactions $k_R \to \infty$ are shown in (a)
and for finite-ranged interactions with $k_R = k_0$ are shown in (b). 
In both panels the two-body binding energy
is $E_B = 0.01 \varepsilon_0$.
The dotted black lines are $T_c^{\rm up1} = \varepsilon_F/8$,
the dotted green lines are  $T_c^{\rm up0}= \pi \rho_s (T = 0)/2$, the
dot-dashed yellow lines are $T_{c}^{\rm mf}$, the solid red lines
are $T_c^{\theta}$ for $E_c \to \infty$, and the dashed blue lines are
$T_c^{\theta}$ for $E_c = 1.5 \rho_s$. 
}
\label{fig:two}
\end{figure}
%

\section{Results} 
\label{sec:results}

The phase-fluctuation critical temperature $T_c^\theta$ is a much tighter
upper bound to the supremum $T_c^{\rm sup}$ in comparison to 
$T_c^{\rm up1}$ based on ${\widetilde \rho}_{s_1}$, $T_c^{\rm up0}$ based on $\rho_s (T = 0)$,
and the mean-field (saddle-point) critical temperature $T_c^{\rm mf}$
obtained by neglecting phase fluctuations.
The self-consistency relations shown in Eqs.~(\ref{eqn:order-parameter}),
(\ref{eqn:number}) and~(\ref{eqn:critical-temperature}) together with
the renormalization group  flow in Eq.~(\ref{eqn:RG-solution})
determine $\mu$, $\vert \Delta \vert$ and $T_c^{\theta}$ as
functions of density $n$ in the continuum or filling factor $\nu$ in the square lattice for
given interaction parameters.

In Fig.~\ref{fig:two}, we show $T_c^\theta$, $T_c^{\rm up0}$, $T_c^{\rm up1}$, and $T_c^{\rm mf}$
versus $n$ for 2D continuum systems, where both the cases of zero-ranged and
finite-ranged interaction potentials are considered. We use temperature/energy 
$T_0 = \epsilon_0 = k_0^2/2m$ and density $n_0 = k_0^2/2\pi$ as units, where $k_0$ is
a reference momentum related to the unit cell length $a$ of a crystal ($k_0 = 2\pi/a$) or 
to the laser wavelength $\lambda$ in a cold-atom
system ($k_0 = 2\pi/\lambda$), in which case, $k_0$ $(\varepsilon_0)$ represents
the recoil momentum (energy). We convert the interaction $V_s$ with interaction range
in momentum space $k_R \sim R^{-1}$
into the two-body binding energy $E_B>0$ via the 
two-body bound state relation
\begin{equation}
{1}/{{V}_s}
=
\sum_{\bf k}
{\vert {\Gamma}_s ( {\bf k} ) \vert^2}
/
{\left[ 2 \varepsilon_{\bf k} + E_B \right]}
\end{equation}
to compare more easily the cases of zero- and finite-ranged interactions.
As seen in Figs.~\ref{fig:two}(a) and~\ref{fig:two}(b), 
a comparison between the phase-fluctuation critical temperature $T_c^{\theta}$ 
for $E_c \to \infty$ (solid red line) or for $E_c = 1.5 \rho_s$ (dashed blue line)
and the standard upper bound $T_c^{\rm up1}$ (dotted black line) shows that $T_c^{\rm up1}$ 
fails miserably at intermediate and high densities $n$, 
being practically useless in that regime.
For parabolic bands,  $T_c^{\rm up0}$ (dotted green line) is equal to $T_c^{\rm up1}$ for all
$n$ due to Galilean invariance, and increases linearly with density $n$ as it is
proportional to the Fermi energy $T_c^{\rm up1} = \varepsilon_F/8$ or
particle density $T_c^{\rm up1} = \pi n/8m$.
Furthermore, $T_c^{\rm mf}$ (dot-dashed yellow line) is always larger than  $T_c^{\theta}$,
exceeds $T_c^{\rm up1} = T_c^{\rm up0}$ at lower densities, and is reliable only when
the density is sufficiently large. 

In Fig.~\ref{fig:two}(b),  $T_c^{\rm mf}$ (dot-dashed yellow line) and
$T_c^{\theta}$ for $E_c \to \infty$ (solid red line) or for $E_c = 1.5 \rho_s$ (dashed blue line)
show a distinctive maximum due to the finite-range of the interaction potential.
For contact interactions, $T_c^{\rm mf}$ and  $T_c^{\theta}$
increase monotonically with density over the range shown in Fig.~\ref{fig:two}(a),
while for a finite-ranged interaction potential, they first increase
and then decrease to form a dome structure as function of density,
as seen in Fig.~\ref{fig:two}(b).
The reason for this dome structure is that the symmetry factor $\Gamma_s ({\bf k})$ acts as
a weight function in Eq.~(\ref{eqn:order-parameter}) for the order parameter, thus reducing
the contribution from the independent-particle density of states to pairing
coming from large momenta.
For contact interactions, $\Gamma_s ({\bf k}) = 1$ for all momenta,
but for finite-ranged interactions,
$\Gamma ({\bf k}) \approx (k/k_R)^{-1/2}$ at large momenta $k \gg k_R$, where
$k = \vert {\bf k} \vert$.
Finally, for the single-band 2D continuum,
the sequence of bounds with respect to $T_c^{\rm sup}$ is always
$T_c^{\rm sup} \le T_c^{\theta} \le T_c^{\rm up0} \le T_c^{\rm up1}$.

\begin{figure}[t]
\begin{center}
\includegraphics[width=0.98\linewidth]{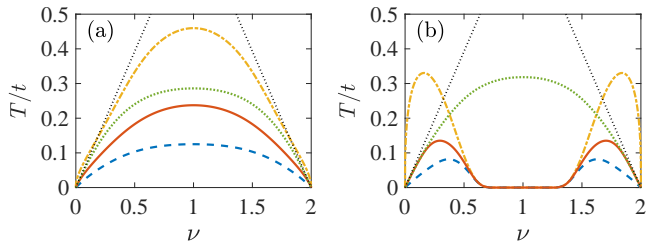}   
\end{center}
\vspace{-6mm}
\caption{
Plots of $T_c$ versus $\nu$. In both panels the interaction parameter is
$V_s/t  = 3$.  In (a) $\Gamma_s ({\bf k}) = 1$ (conventional $s$-wave), and
in (b) $\Gamma_s ({\bf k}) = \cos (k_x a) + \cos (k_y a)$ (extended $s$-wave).
The dotted black lines are the standard upper bound
$T_c^{\rm up1} = t (\pi/4) {\rm min} \{ \nu, (2 - \nu) \}$,
the dotted green lines are $T_c^{\rm up0} = \pi \rho_s (T =0)/2$, 
the dot-dashed yellow lines are $T_{c}^{\rm mf}$,
the solid red lines are $T_c^{\theta}$ for $E_c \to \infty$,
and the dashed blue lines are $T_c^{\theta}$ for $E_c = 1.5 \rho_s$. 
}
\label{fig:three}
\end{figure}

In Fig.~\ref{fig:three}, we show $T_c^\theta$ for $E_c \to \infty$ (solid red line)
and for $E_c = 1.5 \rho_s$ (dashed blue line),
$T_c^{\rm up0}$ (dotted green line),
$T_c^{\rm up1}$ (dotted black line),
and $T_c^{\rm mf}$ (dot-dashed yellow line)
versus $\nu$ for a 2D square lattice with $V_s/t = 3$.
In panel (a) $\Gamma_s ({\bf k}) = 1$ (conventional $s$-wave)
and in (b) $\Gamma_s ({\bf k}) = \cos (k_x a) + \cos (k_y a)$
(extended $s$-wave).
For the conventional $s$-wave case, the choice of $V_s/t= 3$ corresponds to onsite interaction
$U/t = 3$ and nearest neighbor interaction $V/t = 0$ (negative-$U$ Hubbard model),
while for the extended $s$-wave case, the same choice corresponds to $U/t = 0$ and $V/t = 3$
(extended attractive Hubbard model). See the Hamiltonian in
Eq.~(\ref{eqn:extended-attractive-Hubbard-model}).

In the presence of a lattice potential, the independent particle dispersion is no longer parabolic
in momentum space and Galilean invariance is absent. Hence, the superfluid density 
at zero temperature $\rho_s (T = 0)$ is not proportional to the particle density
as it is in the continuum case.  As a result, $T_c^{\rm up0} \ne T_c^{\rm up1}$, and in general
$T_c^{\rm up0} \le T_c^{\rm up1}$ given that ${\tilde{\rho}}_{s_1}$ is always greater than
the superfluid density at zero temperature $\rho_s (T = 0)$.
As shown in Figs.~\ref{fig:three}(a) and~\ref{fig:three}(b),
$T_c^{\rm up0}$ is weakly sensitive to the symmetry of the order parameter, and
$T_c^{\rm up1}$ completely insensitive to the symmetry of the order parameter. 
Both are controlled by the kinetic energy and filling factor $\nu$, that is, by the
independent-particle density of states, and are maximized at half filling $\nu =1$.

In contrast, the qualitative behaviors of $T_c^{\rm mf}$ and $T_c^{\theta}$ are 
dependent on the specific choice of pairing symmetry. 
To calculate $T_c^{\rm mf}$ and  $T_c^{\theta}$, we need
to consider the sensitivity of Eq.~(\ref{eqn:order-parameter}) (order parameter)
to the symmetry factor $\Gamma_s ({\bf k})$, which acts as weight function
in momentum space.
For the isotropic $s$-wave order parameter with $\Gamma_s ({\bf k}) = 1$,
as shown in Fig.~\ref{fig:three}(a), $T_c^{\rm mf}$ and  $T_c^{\theta}$ 
are mostly affected by the variation of the independent-particle density of states
and are peaked at $\nu=1$, producing a single-dome structure.
However, for the anisotropic $s$-wave order parameter with 
$\Gamma_s ({\bf k}) = \cos (k_xa) + \cos (k_ya)$, as shown in Fig.~\ref{fig:three}(b),
$\vert \Gamma_s ({\bf k}) \vert^2$ suppresses the effect of the
independent-particle density of states near half-filling $(\nu = 1)$, and thus 
reduces $T_c^{\rm mf}$ and  $T_c^{\theta}$ in that vicinity, creating a two-dome structure.

As shown in Fig.~\ref{fig:three}, the standard
upper bound $T_c^{\rm up1}$ (dotted black line)
fails miserably in comparison to the phase-fluctuation upper bound $T_c^{\theta}$ 
for $E_c \to \infty$ (solid red line)  or for $E_c = 1.5 \rho_s$ (dashed blue line)
at intermediate fillings $\nu$.  In this regime of intermediate filling factors,
even $T_c^{\rm mf}$ is a better upper bound than $T_c^{\rm up1}$,
however $T_c^{\rm up1}$ is tighter for $\nu \approx 0$ or $\nu \approx 2$.
Notice that $T_c^{\rm up0}$ is always a better upper bound than $T_c^{\rm up1}$, and while in
Fig.~\ref{fig:three}(a) it is not too far above $T_c^{\theta}$, in
Fig.~\ref{fig:three}(b) it overestimates substantially $T_c^{\theta}$
at intermediate $\nu$, where the order parameter modulus $\vert \Delta \vert$ vanishes.
In Fig.~\ref{fig:three}(a), $T_c^{\rm mf}$ is always above all upper bounds for $\nu \approx 0$
and $\nu \approx 2$, but below $T_c^{\rm up1}$ and above $T_c^{\rm up0}$ and $T_c^{\theta}$ at
intermediate $\nu$. While in Fig.~\ref{fig:three}(b), $T_c^{\rm mf}$ is above all upper bounds
for $\nu \approx 0$ or $\nu \approx 2$, but at
intermediate values of $\nu$ it is below $T_c^{\rm up1}$ and $T_c^{\rm up0}$, but always
above $T_c^{\theta}$. In summary, for the single-band 2D square-lattice, the sequence
of bounds with respect to $T_c^{\rm sup}$ is always
$T_c^{\rm sup} \le T_c^{\theta} \le T_c^{\rm up0} \le T_c^{\rm up1}$,
just like in the 2D continuum.

We have seen that to capture the full effects of phase fluctuations it is necessary to include the
renormalization of the chemical potential via the equation of state (number equation).
The renormalization is more important for interactions that are sufficiently strong at fixed density,
or for densities that are sufficiently low at fixed interactions. Thus, one cannot use the
standard textbook approach of pinning the chemical potential to the Fermi energy, and
investigate simply the order parameter and the renormalization of the superfluid density.
We clarify the importance of this statement further in the next section, where we
also discuss effects beyond standard phase fluctuations. 


\section{Beyond Standard Phase Fluctuations}
\label{sec:beyond-standard-phase-fluctuations}

In this section, we discuss two important points regarding the effects of phase fluctuations
and beyond. The first one is how far phase-only fluctuation theories can go in describing the
evolution from the BCS to the Bose regime and how good the phase-fluctuation
critical temperature $T_c^{\theta}$ is as an upper-bound to $T_c^{\rm sup}$. The second point
is how important the effects of modulus fluctuations of the order parameter are in determining
the critical temperature $T_c^{\theta {\rm mod}}$, which includes both phase and modulus
fluctuations of the order parameter.  It is clear that $T_c^{\theta {\rm mod}}$
must be always lower than $T_c^{\theta}$, and thus must be an even
better upper bound to $T_c^{\rm sup}$, that is, $T_c^{\rm sup} \le T_c^{\theta {\rm mod}}
\le T_c^{\theta}$. However, currently, there is no analytical theory that includes
both modulus and phase fluctuations and provides reliable calculations of
$T_c^{\theta {\rm mod}}$, while we present here an analytical theory
that includes the full effects of phase fluctuations, produces physical results
throughout the BCS to Bose evolution, and  gives a much better upper bound
to $T_c^{\rm sup}$ than the conventional upper bounds
$T_c^{\rm up0}$ and $T_c^{\rm up1}$.

\subsection{Comparison to Textbook Phase Fluctuations}
\label{sec:comparison-to-textbook-phase-fluctuations}

In textbook phase-fluctuation theories the chemical potential renormalization is ignored,
that is, the equation of state (number equation) defined in Eq.~(\ref{eqn:number}) is
not included as part of the self-consistent relations.
In such theory, the chemical potential is pinned to the Fermi energy
$\mu = \varepsilon_F$. This  approach is
doomed to fail when interactions are sufficiently strong for fixed density $n$ or
filling factor $\nu$, or when the density $n$ or filling factor $\nu$ are sufficiently low for fixed interactions.
Here, we give an example showing the failure of the textbook phase-fluctuation theory by
contrasting it to our phase-fluctuation theory that interpolates smoothly between the 
BCS and Bose limits. The example we discuss is the 
negative-U Hubbard model with fixed filling factor $\nu$
and changing interaction $U$.

\begin{figure}[t]
\begin{center}
\includegraphics[width=0.98\linewidth]{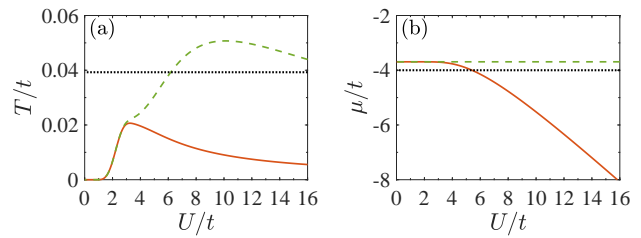}   
\end{center}
\vspace{-6mm}
\caption{
In both panels the filling factor is fixed at $\nu = 0.05$ and $K_c^{-1} = 0.89$
$(E_c =2.22\rho_s)$.
In (a) the phase-fluctuation critical temperatures $T_c^{\theta}$ (solid red line)
including the
renormalization of $\mu$, and $T_c^{\theta}\vert_{\mu = \varepsilon_F}$ (dashed green line)
at $\mu = \varepsilon_F$ are plotted versus $U/t$. The upper bound
$T_c^{\rm up1}/t = \nu \pi/4 = 0.039$ is also shown as the dotted black line.
In (b) the chemical potential $\mu$ (solid red line) obtained including
the equation of state
and $\mu = \varepsilon_F$ (dashed green line) are plotted versus $U/t$.
The value of $\mu$ when it hits the bottom
of the band $\mu/t = -4$ is shown as the dotted black line. 
Notice that $\mu/t \le -4$ for $U/t \ge 5.407$.
}
\label{fig:four}
\end{figure}

First, we look at the regime of $U/t \ll 1$ for filling factor $\nu \le 1$.
In this case, the textbook result (dashed green line) and ours (solid red line)
agree, as can be seen in Fig.~\ref{fig:four}(a). This agreement occurs
because the interaction is sufficiently weak for the chemical potential
to be pinned to the Fermi energy, that is, $\mu \approx \varepsilon_F$
as seen in Fig.~\ref{fig:four}(b).

Next, we look at the regime of $U/t \gg 1$ for fixed filling factor $\nu \le 1$.
The standard textbook phase-fluctuation theory gives $\mu = \varepsilon_F$,
$\vert \Delta \vert = \sqrt{1 - 4 \varepsilon_F^2/U^2}U/2$, and
$\rho_s = (1 - 4\varepsilon_F^2/U^2)t^2/U$. The resulting
phase-only critical temperature is 
\begin{equation}
 T_c^{\theta}\vert_{\mu = \varepsilon_F} 
 = K_c^{-1} \left[1 - 4\varepsilon_F^2/U^2 \right] t^2/U.
\end{equation}
In contrast, for $U/t \gg 1$ and fixed filling factor $(\nu \le 1)$,
the inclusion of the number equation shown in Eq.~(\ref{eqn:number})
gives $\mu =  (\nu -1)U/2 \le 0$,
$\vert \Delta \vert  = \sqrt{\nu (2 - \nu)} U/2$, 
and $\rho_s = \nu (2 - \nu) t^2/U$. This leads to the analytical limit
\begin{equation}
\label{eqn:phase-fluctuation-tc-strong-coupling}
T_c^{\theta} = K_c^{-1} \left[ \nu (2 - \nu) \right]  t^2/U.
\end{equation}
The difference in critical temperatures for these two cases is
$\delta T_c^{\theta} = T_c^{\theta}\vert_{\mu = \varepsilon_F} - T_c^{\theta}$,
where
\begin{equation}
\delta T_c^{\theta} = K_c^{-1} \left[ (1 - \nu)^2 - 4 \varepsilon_F^2/U^2\right] t^2/U.
\end{equation}

Near half-filling $(\nu \approx 1)$, where $\varepsilon_F \approx 0$,
these two phase-fluctuation critical temperatures are close, and thus
$\delta T_c^{\theta} \approx 0$. In this case, the chemical potential lies
near the Fermi energy and two results are approximatelly the same.
However, for low filling factor,  $T_c^{\theta}\vert_{\mu = \varepsilon_F}$ severely
overestimates $T_c^\theta$, because $\mu$ is strongly renormalized away from
$\varepsilon_F$ as $U/t$ grows.
This is clearly seen in Fig.~\ref{fig:four},
where $T_c^{\theta}$ (solid red line) and
$T_c^{\theta}\vert_{\mu = \varepsilon_F}$ (dashed green line)
are plotted versus $U/t$ for $\nu = 0.05$ in panel (a).
Notice that $T_c^{\theta}\vert_{\mu = \varepsilon_F}$ exceeds even the worse
upper bound $T_c^{\rm up1} = t \nu \pi/4$, which here is $T_c^{\rm up1}/t \approx 0.039$,
as indicated by the dotted black line in panel (a). The 
chemical potential is plotted versus $U/t$ for $\nu =0.05$ in panel (b).
The $\mu$ arising from the equation of state in Eq.~(\ref{eqn:number}) is the solid red line
and $\mu = \varepsilon_F$ is the dashed green line.
In both panels, $K_c^{-1} = 0.89$, corresponding to core energy $E_c = 2.22 \rho_s$.
These values are used because they give the correct numerical critical temperature $T_c$
in the large $U/t$ limit $(U/t \gg 1)$,  where a phase-only fluctuation theory is known
to be exact~\cite{erez-2013}, that is, $T_c^{\theta} \to T_c$.
The chemical potential $\mu$ is strongly renormalized as $U/t$ increases,
in particular, for $U/t \ge 5.407$, the chemical potential falls
below the bottom of the band, that is, $\mu/t \le -4$. See panel (b), where the
dotted black line shows $\mu/t = -4$.

Having compared the textbook phase-fluctuation theory results with ours,
and having showed that our approach is much better, producing
physically acceptable results for weak and strong interactions,
we discuss next the effects of modulus fluctuations of the order
parameter.

\subsection{Effects of Modulus Fluctuations}
\label{sec:effects-of-modulus-fluctuations}

Currently, there is no reliable analytical theory that includes the simultaneous
effects of phase and modulus fluctuations of the order parameter, the renormalization
of the chemical potential, and the corresponding  renormalization group equations.
In principle, this theory can be developed, but it is technically challenging.
Nevertheless, a few general statements can be made regarding the inclusion of modulus
fluctuations of the order parameter.

First, the critical temperature $T_c^{\theta {\rm mod}}$
including both phase and modulus fluctuations of the order parameter
can only be smaller than $T_c^{\theta}$, that is,
$T_c^{\theta {\rm mod}} \le T_c^{\theta}$, which
means that the phase-fluctuation critical temperature $T_c^{\theta}$
still serves as better upper bound to $T_c^{\rm sup}$
than $T_c^{\rm up0}$, $T_c^{\rm up1}$ and $T_c^{\rm mf}$. Second, we can
compare our thermodynamic-limit analytical and numerical results for
$T_c^{\theta}$ with the critical temperature $T_c$ obtained by purely
numerical methods~\cite{scalettar-2004, erez-2013} in finite-sized systems, which 
include implicitly the combined effects of phase and modulus fluctuations of the
order parameter.
In comparing our thermodynamic-limit results with numerical calculations
in finite-sized systems,
it is very important to ensure that finite-size effects are negligible. When this is the case, our
results for $T_c^{\theta}$ should always lie above the numerical $T_c$. 
Third, modulus (and phase)  fluctuations of the order parameter also occur
in inhomogeneous systems with disorder, where it has been experimentally found
that the superfluid density is directly connected to the average of the order parameter modulus
when the system is close to a superfluid-insulator transition~\cite{yong-2013}.
Disorder effects may reduce the phase-fluctuation critical temperature $T_c^{\theta}$,
however we do not address the effects of disorder in this paper.
Fourth, modulus fluctuations may further renormalize the superfluid density $\rho_s$ and
the compressibility $\kappa_s$, specially in the Bose regime, where they lead to
small logarithmic corrections to $T_c^{\theta}$ due to residual boson-boson
interactions in the continuum~\cite{fisher-1988}.

\begin{figure}[t]
  \begin{center}
 \vspace{3mm}   
\includegraphics[width=0.98\linewidth]{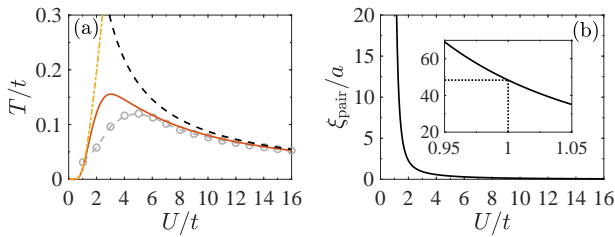}   
\end{center}
\vspace{-6mm}
\caption{
In (a), we plot critical temperatures $T_c/t$ vs. $U/t$ at filling factor $\nu = 0.875$.
The phase-fluctuation critical temperature $T_c^{\theta}$ (solid red line)
is compared to the numerical $T_c$ (gray circles with dashed gray line)
for $20 \times 20$ lattice sites~\cite{erez-2013}.
The core energy is $E_c = 2.22 \rho_s$ and $K_c^{-1} = 0.89$ are the same as in
Ref.~\cite{erez-2013}. The dot-dashed yellow line is the mean-field critical temperature
$T_c^{\rm mf}$ and the dashed black line is the analytical result
$T_c^{\theta} (U/t \gg 1) = K_c^{-1} \left[ \nu (2 - \nu) \right] t^2/U$.
In (b), we plot the ratio $\xi_{\rm pair}/a$ vs. $U/t$
to understand when the numerical results for a $20 \times 20$
lattice~\cite{erez-2013} are reliable.
The inset shows that
$\xi_{\rm pair}/a = 48.344$
at $U/t = 1$, demonstrating that the numerical data at $U/t = 1$ using
a $20 \times 20$ lattice is not reliable.
}
\label{fig:five}
\end{figure}

In all cases discussed above, the only physically acceptable outcome is that the inclusion
of modulus fluctuations of the order parameter must reduce
$T_c^{\theta}$ to $T_c^{\theta {\rm mod}}$. However, in the
absence of a thermodynamic-limit theory that includes both phase
and modulus fluctuations, we can use a solid thermodynamic-limit theory that
includes the effects of phase fluctuations to give a reliable upper bound to $T_c^{\rm sup}$.
To show that our phase fluctuation theory provides a much tighter upper bound $T_c^{\theta}$
than $T_c^{\rm up0}$, $T_c^{\rm up1}$ and $T_c^{\rm mf}$, we compare in Fig.~\ref{fig:five}
our thermodynamic-limit results for the negative-U Hubbard model in 2D 
and numerical work for the same model using a finite-sized lattice with
$20 \times 20$ sites~\cite{erez-2013}.

In Fig.~\ref{fig:five}(a),  we show $T_c ^{\theta}$ (solid red line) and the numerical
$T_c$ (gray circles with dashed gray line) for finite-sized lattices~\cite{erez-2013}
versus interaction strength $U$.  The system described is the 2D negative-$U$
Hubbard model at filling factor $\nu = 0.875$, for which numerical
data is available~\cite{erez-2013} when  $U/t \ge 1$.
The core energy $E_c =2.22\rho_s$ and $K_c^{-1} = 0.89$ are the same values
used in the numerical work~\cite{erez-2013}, since for $U/t \gg 1$ a phase-only
fluctuation theory is sufficient to give the critical temperature, that is,
$T_c^{\theta} \to T_c$.
Notice that our $T_c^{\theta}$ reproduces exactly the known limits for $U/t \ll 1$ given
by $T_c^{\rm mf}$ (dot-dashed yellow line) and for $U/t \gg 1$ given by
$T_c^\theta (U/t \gg 1)$ (dashed black line), and interpolates
between them producing a distinctive maximum at $U/t = 3.04$ with $T_c^{\theta}/t = 0.156$.
We emphasize that $T_c^{\theta}$ agrees with the numerical $T_c$ for $U/t \gg 1$,
where modulus fluctuations are negligible~\cite{erez-2013}. The analytical limit
$T_c^\theta (U/t \gg 1)$ (dashed black line) is that of
Eq.~(\ref{eqn:phase-fluctuation-tc-strong-coupling}).

When $U/t \gg 1$, finite-size effects of the $20 \times 20$ lattice,
used in Ref.~\cite{erez-2013},
are not important, because Cooper pairs are much smaller than the lattice
size, that is, the ratio $\xi_{\rm pair}/a \ll 20$. However, when $U/t \ll 1$ finite size effects
are important because the ratio $\xi_{\rm pair}/a \gg 20$. This means that there
is a characteristic value $U_c/t$ below which the numerical $T_c$
of Ref.~\cite{erez-2013} cannot be trusted and this is why their data
does not go all the way to $U/t = 0$. To determine the importance of finite-size effects
we calculate the pair size~\cite{duncan-2000}
\begin{equation}
\xi_{\rm pair} = 
\left[
  \frac{\sum_{\bf k} \varphi_{\bf k}^* (- \nabla_{\bf k}^2) \varphi_{\bf k}}{\sum_{\bf k}
    \varphi_{\bf k}^* \varphi_{\bf k}}
  \right]^{1/2},
\end{equation}
where $\varphi_{\bf k} = \Delta_{\bf k}/2E_{\bf k}$ is the pair wavefunction, 
$E_{\bf k} = \sqrt{ \xi_{\bf k}^2 + \vert \Delta_{\bf k} \vert^2}$ is the quasiparticle
excitation energy, 
$\Delta_{\bf k} = \Delta \Gamma_s ({\bf k})$ is the order parameter with $\Gamma_s ({\bf k}) = 1$
(conventional s-wave) and $\xi_{\bf k} = \varepsilon_{\bf k} - \mu$ with
$\varepsilon_{\bf k} = -2t \left[ \cos (k_x a) + \cos (k_y a) \right]$.
The ratio $\xi_{\rm pair}/a$ versus
$U/t$, where $a$ is the lattice spacing, is shown in Fig.~\ref{fig:five}(b).
For $U/t \le 1.153$,
the pair size $\xi_{\rm pair}/a \ge 20$, and
finite size effects matter when using a $20 \times 20$ lattice.
This implies that the numerical data~\cite{erez-2013}
at $U/t = 1$, shown in Fig.~\ref{fig:five}(a), cannot be trusted
because $\xi_{\rm pair}/a = 48.344$
is larger than the linear lattice dimension of 20.
This analysis shows that our phase-fluctuation critical
temperature $T_c^{\theta}$ always lies above the numerical results~\cite{erez-2013}
when finite size effects do not matter. For completeness, we mention briefly that the
standard upper bound $T_c^{\rm up1}= t \nu \pi/4$ at $\nu = 0.875$ gives
$T_c^{\rm up1}/t = 0.687$ which is more than 4 times larger than the peak
in $T_c^{\theta}$ (solid red line) in Fig.~\ref{fig:five}(a).

From the comparison above, we learned several things.
First, the numerical $T_c$ in a $20 \times 20$ lattice~\cite{erez-2013}
should always lie below our thermodynamic phase-fluctuation $T_c^{\theta}$
when finite size effects do not matter. This criterion led us to catch a reliability issue
at $U/t = 1$ for the numerical $T_c$ in a $20 \times 20$ lattice.
Second, the differences between $T_c^{\theta}$ and the numerical $T_c$
in a $20 \times 20$ lattice~\cite{erez-2013} can be attributed to modulus fluctuations of
the order parameter when finite size effects are not important. This follows from the
general requirement that $T_c^{\theta {\rm mod}} \le T_c^{\theta}$.
Third, the effects of modulus fluctuations do not seem to be very substantial for the
parameters used, but
they are clearly seen in Fig.~\ref{fig:five}(a) for the region of $2 < U/t < 5$.
Fourth, the conventional upper bound $T_c^{\rm up1} = t \nu \pi/4$ for $\nu = 0.875$
is more than 4 times larger than the peak in $T_c^{\theta}$ (solid red line)
seen in Fig.~\ref{fig:five}(a), reassuring that indeed $T_c^{\theta}$ is always
a better upper bound to $T_c^{\rm sup}$ than
$T_c^{\rm up0}$ and $T_c^{\rm up1}$.

Lastly, we would like to remark that quantum Monte Carlo
(QMC) data~\cite{scalettar-2004}
on $T_c$ versus $\nu$ for the attractive Hubbard model,
a case illustrated in Fig.~\ref{fig:three}(a)
are always upper-bounded by $T_c^{\theta}$, when finite-size effects are negligible.

\subsection{Connection to Experiments}
\label{sec:connection-to-experiments}

As stated in the introduction, several recent experiments have studied
the critical temperature $T_c$ of  2D superconductors as a function of
carrier density $n$ or filling factor $\nu$ for various
materials~\cite{herrero-2018, herrero-2021, iwasa-2018, iwasa-2021, shibauchi-2021},
have raised the issue of the existence of an upper bound on $T_c$, and
have described their results as evolving from the BCS to the Bose regime.

Our renormalization group approach together with the self-consistent relations
for the order parameter, chemical potential and the Nelson-Kosterlitz condition
provide a solid calculation
of the phase-fluctuation critical temperature $T_c^{\theta}$ based on the
Berezinskii-Kosterlitz-Thouless mechanism of vortex-antivortex unbinding.
For one-band superconductors and superfluids, our work addresses the issue
of tighter upper bounds for the critical temperature, transcends the
standard textbook phase fluctuation theory and can be applied throughout
the BCS-Bose evolution in 2D. 

We note, that the renormalized superfluid density $\rho_s^{R}$ is the
physical superfluid density which is obtained via penetration depth
measurements~\cite{uemura-1989, uemura-1991}.
Thus, for 2D superconductors, $T_c^{\theta}$ can be extracted through
the Kosterlitz-Nelson relation in Eq.~(\ref{eqn:critical-temperature}) and compared to
the experimental critical temperature $T_c$ obtained from resistivity measurements.
Since the experimental $T_c$ includes phase and modulus fluctuations of the order parameter,
we must have $T_c \le T_c^{\theta}$, that is, $T_c^{\theta}$ is an upper bound for $T_c$.
If the BKT mechanism fully applies, then $T_c = T_c^{\theta}$. However, if
$T_c  > T_c^{\theta}$, it is clear that a non-BKT mechanism must be invoked. 

\section{Conclusions}
\label{sec:conclusions}

We investigated tighter upper bounds on the critical temperature of two-dimensional (2D)
superconductors and superfluids with a single parabolic (cosinusoidal) band in the continuum
(square lattice). Using the renormalization group, we obtained the phase-only fluctuation
critical temperature $T_c^{\theta}$  as the best upper bound for
the supremum $T_c^{\rm sup}$ within the Berezinksii-Kosterlitz-Thouless (BKT)
vortex-antivortex binding mechanism. We showed
that standard upper bounds which are independent of interactions and order parameter
symmetry are only useful at extremely low carrier density and practically useless anywhere
else. We compared our results with numerical work on finite-sized lattices, and showed
explicitly that the phase-only critical temperature $T_c^{\theta}$ is indeed a better upper
bound than standard ones. Furthermore, we discussed that modulus fluctuations of the order
parameter can only produce a critical temperature that is lower than $T_c^{\theta}$.
However, in  the absence of a reliable thermodynamic analytical theory that includes
both modulus and phase fluctuations, $T_c^{\theta}$ provides the best known
upper bound without resorting to numerical work on finite-sized systems.
Finally, our results have important implications for the measurements of $T_c$ in 
2D one-band superconductors with non-retarded interactions and without disorder:
If $T_c$, measured resistively, exceeds $T_c^{\theta}$, obtained from penetration
depth measurements, then a non-BKT mechanism must be invoked.

\acknowledgments{
We thank the National Key R$\&$D Program of China (Grant 2018YFA0306501),
the National Natural Science Foundation of China (Grants 12074428, 92265208)
and the Beijing Natural Science Foundation (Grant Z180013)
for financial support.}



\begin{thebibliography}{2}

\bibitem{herrero-2018}
Y. Cao, V. Fatemi, S. Fang, K. Watanabe, T. Taniguchi, E. Kaxiras, and P. Jarillo-Herrero,
Unconventional superconductivity in magic-angle graphene superlattices,
Nature {\bf 556}, 43 (2018).
  
\bibitem{herrero-2021}
J. M. Park, Y. Cao, K. Watanabe, T. Taniguchi, and P. Jarillo-Herrero,
Tunable strongly coupled superconductivity in magic-angle twisted trilayer graphene,
Nature {\bf 590}, 249 (2021).

\bibitem{iwasa-2018}
Y. Nakagawa, Y. Saito, T. Nojima, K. Inumaru, S. Yamanaka, Y. Kasahara, and Y. Iwasa,
Gate-controlled low carrier density superconductors: Toward the two-dimensional
BCS-BEC crossover,
Phys. Rev. B {\bf 98}, 064512 (2018).

\bibitem{iwasa-2021}
Y. Nakagawa, Y. Kasahara, T. Nomoto, R. Arita, T. Nojima, and Y. Iwasa, 
Gate-controlled BCS-BEC crossover in a two-dimensional superconductor,
Science {\bf 372}, 190 (2021).

\bibitem{shibauchi-2021}
Y. Mizukami, M. Haze, O. Tanaka, K. Matsuura, D. Sano,
J. B\"{o}ker, I. Eremin, S. Kasahara, Y. Matsuda, and T. Shibauchi,
Thermodynamics of transition to BCS-BEC crossover superconductivity
in ${\rm FeSe}_{1-x}{\rm S}_x$,
arXiv:2105.00739v1 (2021).

\bibitem{botelho-2006}
S. S. Botelho and C. A. R. S\'{a} de Melo,
Vortex-Antivortex Lattice in Ultracold Fermionic Gases,
Phys. Rev. Lett.  {\bf 96}, 040404 (2006).

\bibitem{sharapov-1999}
A similar expression relating $T_c$ to $\varepsilon_F$ at small densities was obtained by 
V. P. Gusynin, V. M. Loktev and S. G. Sharapov,
Pseudogap phase formation in the crossover from Bose–Einstein condensation
to BCS superconductivity, JETP {\bf 88},  685 (1999). However, these authors did
not recognize that this relation corresponds to an upper bound to $T_c$ for all densities,
independent of interaction strength.

\bibitem{randeria-2019}
T. Hazra, N. Verma, and M. Randeria,
Bounds on the Superconducting Transition Temperature:
Applications to Twisted Bilayer Graphene and Cold Atoms,  
Phys.  Rev. X {\bf 9}, 031049 (2019).

\bibitem{kivelson-2021}
J. S. Hofmann, D. Chowdhury, S. A. Kivelson, and E. Berg,
Heuristic bounds on superconductivity and how to exceed them,
npj Quantum Mater. {\bf 7}, 83 (2022). See also: arXiv:2105.09322v2 (2021).

\bibitem{kivelson-2018}
I. Esterlis, S. Kivelson, and D. Scalapino,
A bound on the superconducting transition temperature,
npj Quantum Mater. {\bf 3}, 59 (2018).

\bibitem{kivelson-1995}
V. J. Emery and S. A. Kivelson, 
Importance of phase fluctuations in superconductors
with small superfluid density,
Nature {\bf 374}, 434 (1995).

\bibitem{uemura-1989}
Y. J. Uemura, G. M. Luke,  B. J. Sternlieb, J. H. Brewer, J. F. Carolan, W. N. Hardy,
R. Kadono, J. R. Kempton, R. F. Kiefl, S. R. Kreitzman, P. Mulhern, T. M. Riseman,
D. Ll. Williams, B. X. Yang, S. Uchida, H. Takagi, J. Gopalakrishnan, A. W. Sleight,
M. A. Subramanian, C. L. Chien, M. Z. Cieplak, Gang Xiao, V. Y. Lee, B. W. Statt,
C. E. Stronach, W. J. Kossler, and X. H. Yu,
Universal Correlations between T, and $n_s /m^*$ (carrier density over effective mass)
in high-$T_c$ cuprate superconductors,
Phys. Rev. Lett. {\bf 62},  2317  (1989).

\bibitem{uemura-1991}  
Y. J. Uemura,  L. P. Le,  G. M. Luke,  B. J. Sternlieb,  W. D. Wu,  J. H. Brewer, T. M.
Riseman, C. L. Seaman, M. B. Maple, M. Ishikawa, D. G. Hinks, J. D. Jorgensen, G.
Saito, and H. Yamochi,
Basic similarities among cuprate, bismuthate, organic, Chevrel-phase, and heavy-fermion
superconductors shown by penetration-depth measurements,
Phys. Rev. Lett. {\bf 66},  2665 (1991).

\bibitem{footnote-0}
To illustrate the point with an analogy, consider the distance between the Earth
and the Moon,  which is certainly smaller than the distance between the Earth and the Sun.
However, using the Earth-Sun distance as an upper bound for the Earth-Moon
distance is certainly an overkill, as the Moon apogee is the least upper bound (supremum)
for the Earth-Moon distance

\bibitem{berezinskii-1970}
V. L. Berezinskii,
Destruction of long-range order in one-dimensional and two-dimensional systems having
a continuous symmetry group : I. Classical systems,   
Sov. Phys. JETP {\bf 32}, 493 (1970).

\bibitem{kosterlitz-thouless-1972}
J. M. Kosterlitz and D. Thouless,
Long-range order and metastability in two dimensional solids and superfluids: Application of
dislocation theory.
J. Phys. C {\bf 5}, L124 (1972).

\bibitem{duncan-2000}
R. D. Duncan and C. A. R. S{\' a} de Melo, 
Thermodynamic properties in the evolution from BCS to Bose-Einstein condensation
for a d-wave superconductor at low temperatures,
Phys. Rev. B {\bf 62}, 9675 (2000).

\bibitem{annett-1990}
J. F. Annett,
Symmetry of the order parameter for high-temperature superconductivity,
Adv. Phys. {\bf 39}, 83  (1990).

\bibitem{iskin-2005}
M. Iskin and C. A. R. S{\' a} de Melo,
Superfluidity of p-wave and $s$-wave atomic Fermi gases in optical lattices
Phys. Rev. B {\bf 72}, 224513 (2005).

  
\bibitem{nelson-kosterlitz-1977}
D. R. Nelson and J. M. Kosterlitz,
Universal jump in the superfluid density of two-dimensional superfluids,
Phys. Rev. Lett. {\bf 39}, 1201 (1977).  

\bibitem{kosterlitz-1974}
J. M. Kosterlitz,
The critical properties of the two-dimensional xy model,  
J. Phys. C: Solid State Phys. {\bf 7}, 1046 (1974).

\bibitem{nelson-1977}
J.  V. Jos\'e, L. P. Kadanoff, S. Kirkpatrick, and D. R. Nelson,
Renormalization, vortices, and symmetry-breaking perturbations in the
two-dimensional planar model,
Phys. Rev. B {\bf 16}, 1217 (1977); Erratum: Phys. Rev. B {\bf 17}, 1477 (1978).

\bibitem{chaikin-1995}
P. M. Chaikin and T. C. Lubensky,
{\it Principles of condensed matter physics},
Cambridge University Press, 1995.

\bibitem{ferrell-1958}
R. A. Ferrell and R. E. Glover,
Conductivity of superconducting films: A sum rule,
Phys. Rev. {\bf 109}, 1398 (1958).
  
\bibitem{tinkham-1959}
M. Tinkham and R. A. Ferrell,
Determination of the superconducting skin depth from the energy gap and sum rule,
Phys. Rev. Lett. {\bf 2}, 331 (1959).

\bibitem{tinkham-1975}
M. Tinkham,
Introduction to Superconductivity,
McGraw-Hill, New York, 1975.

\bibitem{kubo-1957}
R. Kubo,
Statistical-Mechanical Theory of Irreversible Processes.
I. General Theory and Simple Applications to Magnetic and Conduction Problems, 
J. Phys. Soc. Jpn. {\bf 12}, 570 (1957).  

\bibitem{devreese-2014}
J. P. A. Devreese, J. Tempere, and C. A. R. S{\'a} de Melo,
Effects of Spin-Orbit Coupling on the Berezinskii-Kosterlitz-Thouless Transition
and the Vortex-Antivortex Structure in Two-Dimensional Fermi Gases,
Phys. Rev. Lett. {\bf 113}, 165304 (2014).

\bibitem{devreese-2015}
J. P. A. Devreese, J. Tempere, and C. A. R. S{\'a} de Melo,
Quantum phase transitions and Berezinskii-Kosterlitz-Thouless temperature
in a two-dimensional spin-orbit-coupled Fermi gas,
Phys. Rev. A {\bf 92}, 043618 (2015).

\bibitem{hirsch-2000}
J. E. Hirsch and F. Marsiglio,
Optical sum rule violation, superfluid weight, and condensation energy in the cuprates,
Phys. Rev. B {\bf 62}, 15131 (2000).

\bibitem{chubukov-2011}
S. Maiti and A. V. Chubukov,
Optical integral and sum-rule violation in high-$T_c$ superconductors,
Phys. Rev. B {\bf 81}, 245111 (2010).
 
\bibitem{scalettar-2004}
T. Paiva, R. R. dos Santos, R. T. Scalettar, and P. J. H. Denteneer,
Critical temperature for the two-dimensional attractive Hubbard model,
Phys. Rev. B {\bf 69}, 184501 (2004).

\bibitem{erez-2013}
A. Erez and Y. Meir, Effect of amplitude fluctuations on the Berezinskii-Kosterlitz-Thouless transition
Phys. Rev. B {\bf 88}, 184510 (2013).

\bibitem{yong-2013}
J. Yong, T. R. Lemberger, L. Benfatto, K. Ilin, and M. Siegel, Robustness of the 
Berezinskii-Kosterlitz-Thouless transition in ultrathin NbN films
near the superconductor-insulator transition, Phys. Rev. B {\bf 87}, 184505 (2013).

\bibitem{fisher-1988}
D. S. Fisher, and P. C. Hohenberg,
Dilute Bose gas in two dimensions,
Phys. Rev. B {\bf 37}, 4936 (1988).

\end{thebibliography}
\end{document}